\documentclass[conference]{IEEEtran}
\IEEEoverridecommandlockouts
\usepackage{cite}
\usepackage{amsmath,amssymb,amsfonts}
\usepackage{algorithmic}
\usepackage{graphicx}
\usepackage{caption} 
\usepackage{textcomp}
\usepackage{xcolor}
\usepackage{subfigure}
\usepackage{multirow} 

\def\myModel{{XSema}} 

\def\BibTeX{{\rm B\kern-.05em{\sc i\kern-.025em b}\kern-.08em
    T\kern-.1667em\lower.7ex\hbox{E}\kern-.125emX}}
\begin{document}

\title{\myModel: A Novel Framework for Semantic Extraction of Cross-chain Transactions

\thanks{* Corresponding author.}
}

\author{\IEEEauthorblockN{Ziye Zheng}
\IEEEauthorblockA{\textit{School of Computer Science and Engineering} \\
\textit{Sun Yat-sen University}\\
Guangzhou, China \\
zhengzy39@mail2.sysu.edu.cn}
\and
\IEEEauthorblockN{Jiajing Wu}
\IEEEauthorblockA{\textit{School of Software Engineering} \\
\textit{Sun Yat-sen University}\\
Zhuhai, China \\
wujiajing@mail.sysu.edu.cn}
\and
\IEEEauthorblockN{Dan Lin*}
\IEEEauthorblockA{\textit{School of Software Engineering} \\
\textit{Sun Yat-sen University}\\
Zhuhai, China  \\
lind8@mail2.sysu.edu.cn}
\and
\IEEEauthorblockN{Quanzhong Li}
\IEEEauthorblockA{\textit{School of Computer Science and Engineering} \\
\textit{Sun Yat-sen University}\\
Guangzhou, China \\
liquanzh@mail.sysu.edu.cn}
\and
\IEEEauthorblockN{Na Ruan}
\IEEEauthorblockA{\textit{School of Computer Science and Engineering} \\
\textit{Shanghai Jiaotong University}\\
Shanghai, China \\
naruan@sjtu.edu.cn}
}
\maketitle

\begin{abstract}
As the number of blockchain platforms continues to grow, the independence of these networks poses challenges for transferring assets and information across chains. Cross-chain bridge technology has emerged to address this issue, establishing communication protocols to facilitate cross-chain interaction of assets and information, thereby enhancing user experience. However, the complexity of cross-chain transactions increases the difficulty of security regulation, rendering traditional single-chain detection methods inadequate for cross-chain scenarios. Therefore, understanding cross-chain transaction semantics is crucial, as it forms the foundation for cross-chain security detection tasks. Although there are existing methods for extracting transaction semantics specifically for single chains, these approaches often overlook the unique characteristics of cross-chain scenarios, limiting their applicability. This paper introduces \myModel, a novel cross-chain semantic extraction framework grounded in asset transfer and message-passing, designed specifically for cross-chain contexts. Experimental results demonstrate that \myModel~effectively distinguishes between cross-chain and non-cross-chain transactions, surpassing existing methods by over {9\%} for the generality metric and over {10\%} for the generalization metric. Furthermore, we analyze the underlying asset transfer patterns and message-passing event logs associated with cross-chain transactions. We offer new insights into the coexistence of multiple blockchains and the cross-chain ecosystem.

\end{abstract}


\section{Introduction}


Blockchain technology, as a decentralized distributed ledger, has garnered significant attention and investment from governments, financial institutions, technology sectors, and capital markets due to its transparency, security, and immutability~\cite{zheng2017overview}. In recent years, the increasing demand for specific applications and continuous technological advancements have led to a proliferation of blockchain platforms, gradually creating an ecosystem characterized by the coexistence of multiple chains. However, the inherent ``independence" of blockchain results in a lack of effective communication channels between different chains, leading to an ``island" effect that complicates the transfer of assets and information across multiple chains. This situation somewhat undermines the original goal of decentralized interconnectivity~\cite{ou2022overview}.


In September 2016, Vitalik Buterin, the founder of Ethereum, systematically addressed the issue of blockchain interoperability for the first time at the R3CEV conference, marking the formal introduction of the cross-chain concept into academic discourse~\cite{swan2015blockchain,ou2022overview}. As user expectations for seamless transfers of assets and information continue to rise, cross-chain interoperability has become an urgent necessity within the blockchain domain, leading to the rapid development of cross-chain bridge technology~\cite{belchior2021survey}. A cross-chain bridge is a technical architecture that facilitates the interaction of assets and information between different blockchains by establishing communication protocols and trust verification mechanisms. This allows users to initiate transactions on a source chain that can impact the state of a destination chain. Based on their trust and verification models, cross-chain bridges can be categorized into centralized cross-chain bridges (CeFi bridges) and decentralized cross-chain bridges (DeFi bridges)~\cite{trustspectrum2022}. CeFi bridges rely on centralized entities to manage digital assets, while DeFi bridges utilize smart contracts to provide decentralized functionality and transparency. Core technologies such as sidechains, relay chains, and hash locking have addressed the challenges of cross-chain interaction to varying degrees, facilitating the free flow of assets and information~\cite{lee2023sok}.

The rapid development of cross-chain bridge technology has significantly enhanced the interoperability and user experience of the blockchain ecosystem. However, it has also introduced new complexities in transaction semantics within the Web3 environment, posing security regulatory challenges. In October 2022, Elliptic reported an escalation in cross-chain crime threats~\cite{elliptic2022}, revealing that approximately {\$4.1} billion in illegally laundered assets flowed through services such as cross-chain bridges, with this figure rising to {\$7} billion a year later~\cite{elliptic2023}. This indicates that cross-chain technology has become a new tool for criminals. Cross-chain operations not only complicate regulation but also enhance the anonymity of transactions. This increased anonymity may render effective detection methods for financial fraud~\cite{lin2024riskprop}, hacking~\cite{wu2024dappfl}, mixings~\cite{wu2021detecting}, and money laundering~\cite{lin2024denseflow,wu2023towards} that work in a single-chain environment ineffective in a cross-chain context. To address these challenges, cross-chain semantic extraction has become crucial for effectively monitoring and tracking security risks, primarily in two areas: First, precise identification of cross-chain transaction semantics can assist regulatory bodies in distinguishing between non-cross-chain and cross-chain transactions, thereby establishing connections between inter-chain transactions and creating a comprehensive monitoring framework to maintain the stability of the blockchain ecosystem. Second, accurate extraction of cross-chain transaction semantics aids in adapting single-chain monitoring methods to multi-chain scenarios~\cite {wu2023financial}, enhancing overall security and improving the adaptability and effectiveness of regulatory mechanisms.



Semantic extraction is not unique to blockchain; it is widely employed in various data analysis contexts~\cite{liu2024trustgo}. Transaction semantic extraction is a specific application of this concept, focusing on extracting information from transaction data. While this task is not new, the increasing diversity of blockchain scenarios has limited the applicability of existing methods in cross-chain environments. Many blockchain semantic extraction techniques~\cite{wang2021towards,zhou2020ever,cai2023ponzi,wu2021defiranger,xia2021trade} primarily target specific financial security issues, making direct migration to cross-chain contexts challenging. While some generalized methods~\cite{wu2023know} exhibit a degree of portability in cross-chain tasks, they often fail to adequately address the unique challenges posed by cross-chain applications, resulting in reduced effectiveness. Consequently, current research remains insufficient in its consideration of cross-chain scenarios and faces several major challenges:
\begin{itemize}
    \item \textbf{C1: Diversity of Cross-Chain Mechanisms.} Different cross-chain bridges employ various mechanisms and generate distinct event logs, posing a significant challenge in developing a universal method for cross-chain transaction identification.
    \item \textbf{C2: Proliferation of Cross-Chain Bridge Products.} As cross-chain technology evolves, the number, types, and complexity of cross-chain bridges are expected to increase significantly. Effectively identifying transactions initiated through new cross-chain bridges presents an additional major challenge.
\end{itemize}

\begin{figure}[htbp]
    \centering
    \includegraphics[width=1\linewidth]{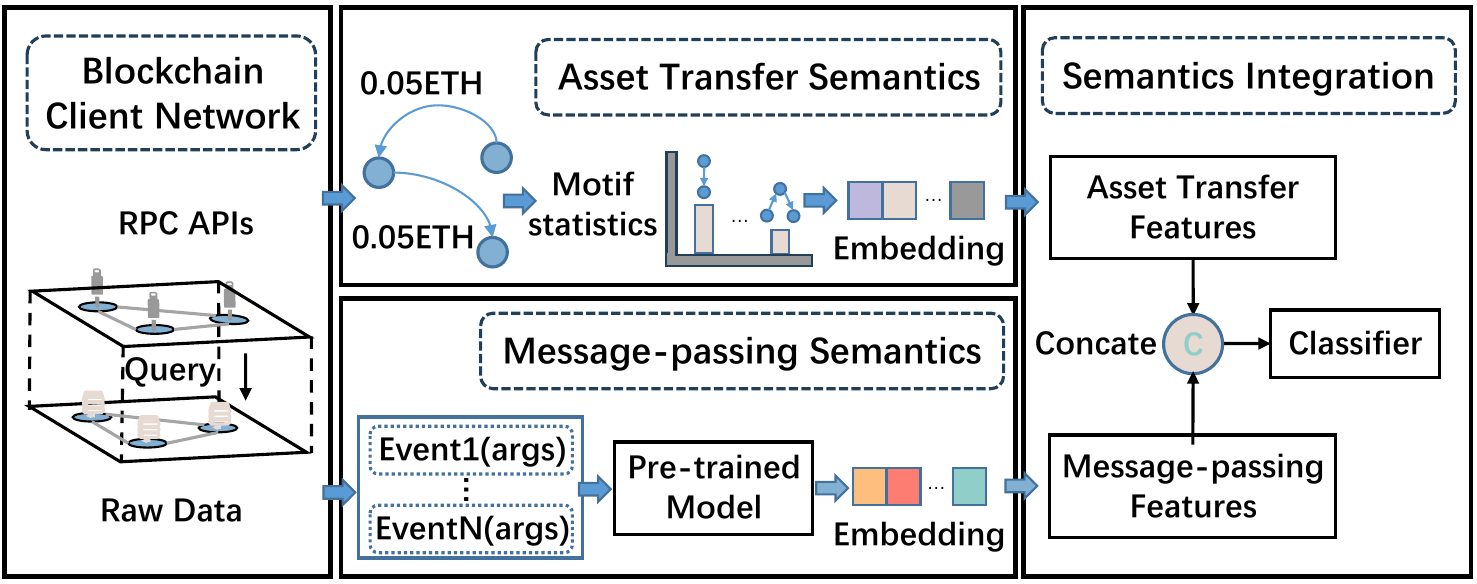}
    \caption{Overview of \myModel. The framework inputs the asset transfer graph and event log text, producing transaction types that include non-cross-chain, deposit, and withdrawal transactions.}
    \label{fig:framework}
\end{figure}

In this paper, we propose a semantic extraction framework named \myModel~to address the problem of semantic extraction for cross-chain transactions. First, we analyze the transaction execution process of cross-chain bridges, identifying two core elements involved: asset transfer and message passing. To tackle \textbf{C1}, we employ a code pre-training model to generalize the event logs, thereby implementing a versatile approach that is not confined to the semantics of specific events. To address \textbf{C2}, we develop an effective method by learning the generalized paradigms underlying cross-chain asset transfer and message-passing mechanisms. The framework of our work is illustrated in Fig.~\ref{fig:framework}.

To evaluate the effectiveness of the \myModel~framework, we collect cross-chain transaction pairs, consisting of deposit transactions from the source chain and withdrawal transactions from the destination chain, across 10 mainstream open-source cross-chain bridges. This results in the creation of the first cross-chain semantic extraction dataset, featuring 11,879 cross-chain transaction pairs and 10,183 non-cross-chain transactions. A detailed description of the transaction pairs will be provided in Section III. Experimental results demonstrate that our framework achieves up to {99.72\%} accuracy in cross-chain semantic extraction for generality while achieving {94.81\%} accuracy for generalizability. Additionally, we analyze the asset transfer patterns and message-passing events associated with the cross-chain transactions identified by \myModel~. Our findings indicate that the distribution of network motifs in cross-chain transactions is more concentrated than that of non-cross-chain transactions, exhibiting unique structural characteristics. Furthermore, the event logs contain specialized terminology related to cross-chain mechanisms, reflecting distinct textual features. In summary, our main contributions include:
\begin{itemize}
    \item \textbf{Data Collection.} We construct the first cross-chain semantic extraction dataset, comprising 11,879 cross-chain transaction pairs and 10,183 on-chain transactions from 10 major cross-chain bridges.
    \item \textbf{Framework Development.} We propose a versatile framework with high generality and generalizability for the semantic extraction of cross-chain transactions, enabling effective distinction between cross-chain deposit transactions, cross-chain withdrawal transactions, and non-cross-chain transactions. 
    \item \textbf{Pattern Analysis.} We analyze the asset transfer patterns and message-passing events underlying cross-chain transactions, providing new insights for the study of cross-chain transaction semantics.
\end{itemize}

The remainder of this paper is organized as follows: Section II reviews related work, while Section III presents our proposed method. Section IV evaluates the effectiveness of our approach through several experiments, and also includes an analysis and summary of the cross-chain semantic patterns.

\section{Related work}

In this section, we examine the research focused on extracting transaction semantics for both schema-specific and generalized patterns within the blockchain context.

\subsection{Schema-Specific Semantic Extraction}
Recent research has focused on understanding transaction semantics to enhance security in specific application contexts. For example,~\cite{wang2021towards,zhou2020ever} have extracted hidden attack transaction semantics, such as fashion-loan and reentrancy, based on predefined transaction-specific patterns established by experts. In the analysis of transaction semantics related to blockchain Ponzi schemes, a novel code representation method~\cite{cai2023ponzi} known as the slice transaction property graph (sTPG) has been proposed. This approach identifies potential Ponzi transaction patterns by mapping the transaction-related semantics of smart contracts to graph structures and leveraging graph neural network techniques. Additionally, DeFiRanger~\cite{wu2021defiranger} models asset transfer relationships as a Currency Flow Tree (CFT) to classify DeFi transaction semantics. Indexing API tools, such as The Graph~\cite{thegraph2022},~\cite{xia2021trade} have also been employed to query specific DeFi semantics, including ``Mint," ``Swap," and ``Burn."

\subsection{Generalized Semantic Extraction}
To enhance the generality of transaction semantic extraction methods,~\cite{wu2023know} have proposed a transaction transfer-based semantic representation utilizing motifs, capable of capturing generic transaction semantic information within real-time transaction data workflows.

Despite these advancements, existing approaches predominantly concentrate on specific security issues or single-chain scenarios, inadequately addressing cross-chain environments' diversity and unique challenges. Consequently, current research remains insufficient regarding the applicability of cross-chain semantic extraction.

\section{Methods}
As discussed in the previous section, research on transaction semantic extraction in cross-chain scenarios remains relatively nascent. Existing methods for transaction semantic extraction exhibit limitations regarding their applicability to cross-chain semantics. To address this issue, this paper proposes a framework for the semantic representation of cross-chain transactions, grounded in asset transfer and message-passing, which we designate as \myModel.

In this section, we will first define the concept of cross-chain transaction semantic extraction. Following this, we will outline the motivation behind the development of the \myModel~framework and detail the implementation of its various modules.

\subsection{Definition}

As previously mentioned, the core function of a cross-chain bridge is to enable users to exchange or transfer digital assets from one blockchain (referred to as the source) to another (referred to as the destination). During the execution of a cross-chain transaction, we observe two critical and spatially independent transactions occurring sequentially: first, a deposit transaction initiated on the source chain, followed by a withdrawal transaction executed on the destination chain. Accurate identification of these two transaction steps is essential; they not only ensure the integrity and transparency of cross-chain transactions but also provide users and regulatory authorities with clear and precise transaction records. Thus, the central task of cross-chain transaction semantic extraction lies in accurately distinguishing between deposit transactions on the source chain, withdrawal transactions on the destination chain, and non-cross-chain transactions.

\textbf{Problem (Cross-Chain Transaction Semantic Extraction)}: Given a transaction $tx_{h}$ with hash value $h$, the corresponding metadata $\mathbb{M}_{h}$ must be retrieved from the chain where it resides. The description of each field in the transaction metadata $\mathbb{M}_{h}$ required by our work is detailed in TABLE~\ref{tab:metadata}. As illustrated in Eq.~\eqref{eq:se_define}, $\mathbb{M}_{h}$ is input into the model to extract the transaction semantics $T_{h}$: whether it is a cross-chain deposit transaction (denoted as $DT$), a cross-chain withdrawal transaction (denoted as $WT$), or a non-cross-chain transactions (denoted as $NT$).

\vspace{-1ex}
\begin{equation}
T_{h}=\myModel(\mathbb{M}_{h})\in\{DT,WT,NT\}\label{eq:se_define}
\end{equation}

\begin{table}
\renewcommand{\arraystretch}{1.6}
  \centering
  \caption{Description of the fields of metadata $\mathbb{M}$.}
  \scalebox{0.78}{
    \begin{tabular}{ll}
    \hline
    Symbol & Description \\ \hline
    $\mathbb{M}_{h}.et$  & External transactions for $tx_{h}$, each including from, to, and value attributes       \\ 
    $\mathbb{M}_{h}.it$  & Internal transactions for $tx_{h}$, each including from, to, and value attributes       \\ 
    $\mathbb{M}_{h}.erc20$  & Erc-20 transactions for $tx_{h}$, each including from, to, and value attributes       \\ 
    $\mathbb{M}_{h}.erc721$  & Erc-721 transactions for $tx_{h}$, each including from, to, and value attributes       \\ 
    $\mathbb{M}_{h}.el$  & Event logs for $tx_{h}$, each including event name attribute       \\ \hline
    \end{tabular}
  }
  \label{tab:metadata}%
\end{table}%

\subsection{Motivation}

\begin{figure}[htbp]
    \centering
    \includegraphics[width=0.95\linewidth]{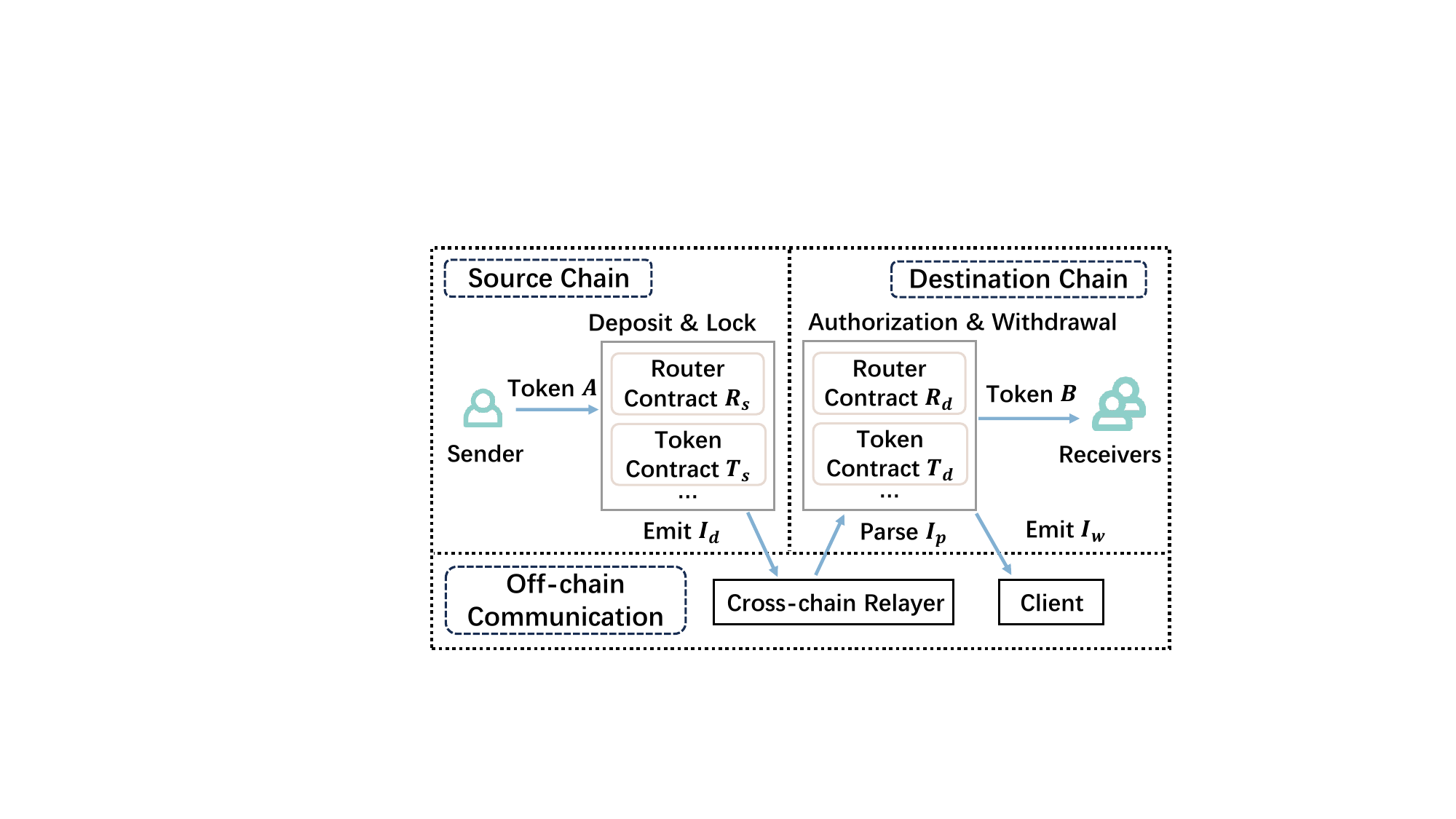}
    \caption{The components of cross-chain bridge.}
    \label{fig:cross-chain}
\end{figure}

Typically, a cross-chain bridge consists of three components: the source chain, the cross-chain relay, and the destination chain. The cross-chain bridge deploys smart contracts on both the source and destination chains, facilitating information exchange between them through the relay, as shown in Fig.~\ref{fig:cross-chain}. Specifically, the execution process of a cross-chain transaction encompasses the following three steps~\cite{liao2024smartaxe}:

\textbf{Asset Deposit on the Source Chain}. Upon receiving an asset exchange request from a sender user, the Router contract $R_{s}$ on the source chain invokes the Token contract $T_{s}$ to lock Token $A$. Subsequently, $R_{s}$ issues a Deposit Event $I_{d}$ as confirmation of the locked asset, which includes detailed deposit information (e.g., type and amount). The user's assets are transferred to the Router contract $R_{s}$.

\textbf{Cross-Chain Communication via Off-Chain Relayers}. Once the Deposit Event $I_{d}$ is issued, the off-chain relay verifies the validity of the deposit on the source chain. If the verification is successful, the relay transmits the informed information $I_{p}$ to the Router contract $R_{d}$ on the destination chain.

\textbf{Asset Withdrawal on the Destination Chain}. Upon receipt of $I_{p}$ at the destination chain, the Router contract $R_{d}$ verifies the various proofs contained within $I_{p}$ to obtain authorization. After successful verification, $R_{d}$ issues a Withdrawal Event $I_{w}$ and invokes the Token contract $T_{d}$ to withdraw the tokens $B$ to the recipient address specified by the sender user.

By analyzing the cross-chain operation flow, we can identify two key elements: asset transfer (i.e., the cross-chain transfer of tokens) and message-passing (facilitated through event logs). In essence, when the cross-chain bridge executes the transfer of user assets, its core mechanism involves the maintenance and synchronization of records related to cross-chain activities. Building upon these two key elements, we develop a framework, \myModel, designed to analyze and identify whether a specific transaction qualifies as a cross-chain transaction. As illustrated in Fig.~\ref{fig:framework}, the framework comprises three modules: asset transfer semantic extraction, message-passing semantic extraction, and semantic integration. 

\addtolength{\topmargin}{0.08in}
\subsection{Asset Transfer Semantics}

In the \myModel~framework we designed, asset transfer semantic extraction primarily focuses on the elements of cross-chain asset transfers. This section explores the invocation relationships between users and contracts that are closely related to asset transfers. Depending on whether the behavior is cross-chain or non-cross-chain, these asset transfer relationships may exhibit unique structural features.

\begin{figure}[htbp]
    \centering
    \includegraphics[width=0.95\linewidth]{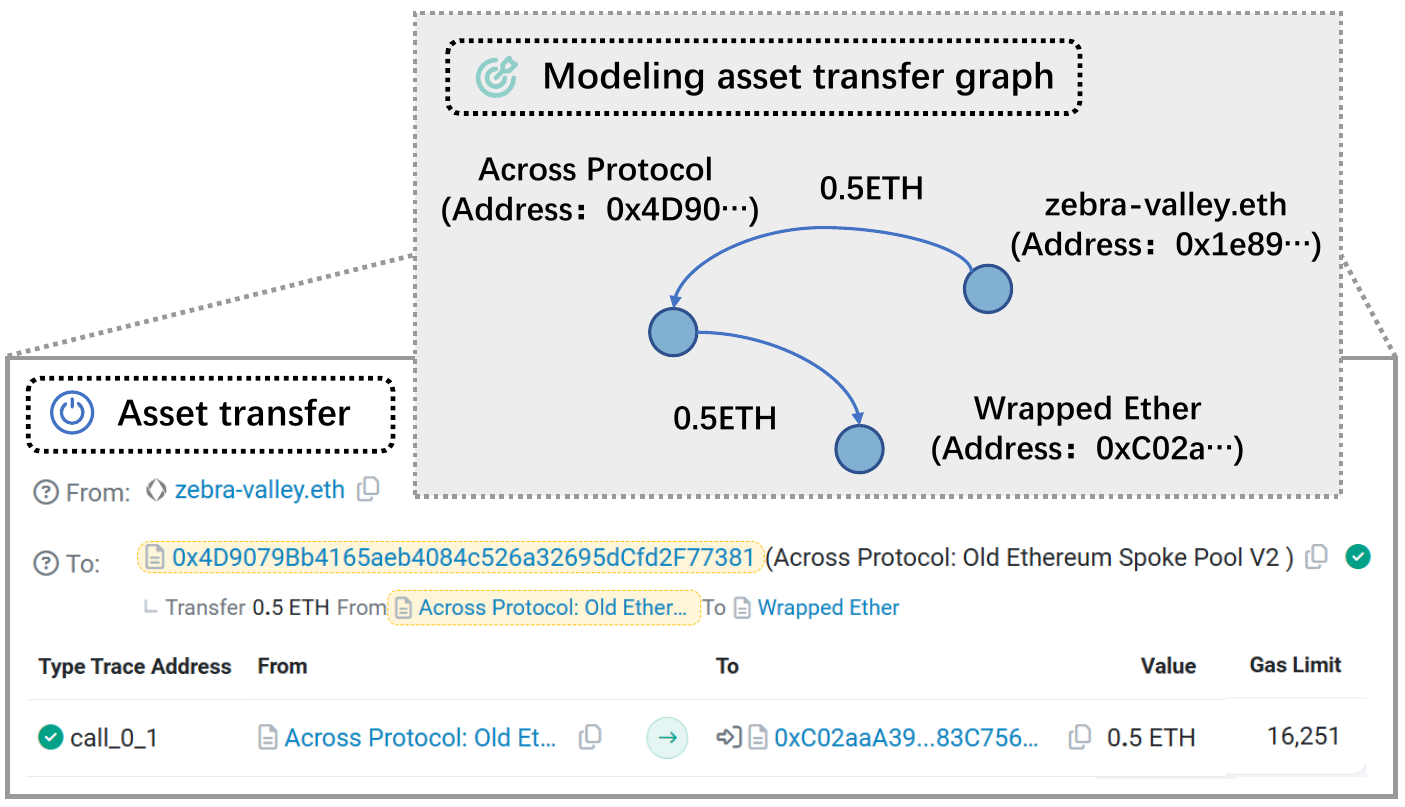}
    \caption{Asset Transfer Graph Modeling.}
    \label{fig:asset}
\end{figure}

\textbf{Asset Transfer Graph Modeling:} We utilize all asset invocation relationships involved in the transaction $tx_{h}$ to construct a corresponding asset transfer graph. Specifically, we gather external transactions, internal transactions, ERC-20 transactions, and ERC-721 transactions related to the transaction through RPC APIs provided by full blockchain nodes. Each asset transfer behavior is represented as edges in the graph, forming a set of edges. Similarly, all asset sender and receiver addresses involved in the transaction are mapped as nodes in the graph, constituting a set of nodes. For example, the asset transfer graph for transaction 0x00f2\footnote{Ethereum transaction hash: 0x00f2f49162f5f31e2419085f37b282a66d5ef1\\076e97fa21d5223fe215cca46b.} in the Ethereum network is shown in Fig.~\ref{fig:asset}. This graph consists of three addresses: {(0x1e89, zebra-valley.eth), (0x4D90, Across Protocol), (0xC02a, Wrapped Ether)} and two asset transfer relationships: {(zebra-valley.eth, Across Protocol, external assets transfer, 0.5 ETH), (Across Protocol, Wrapped Ether, internal assets transfer, 0.5 ETH)}.

\begin{figure}[htbp]
    \centering
    \includegraphics[width=0.95\linewidth]{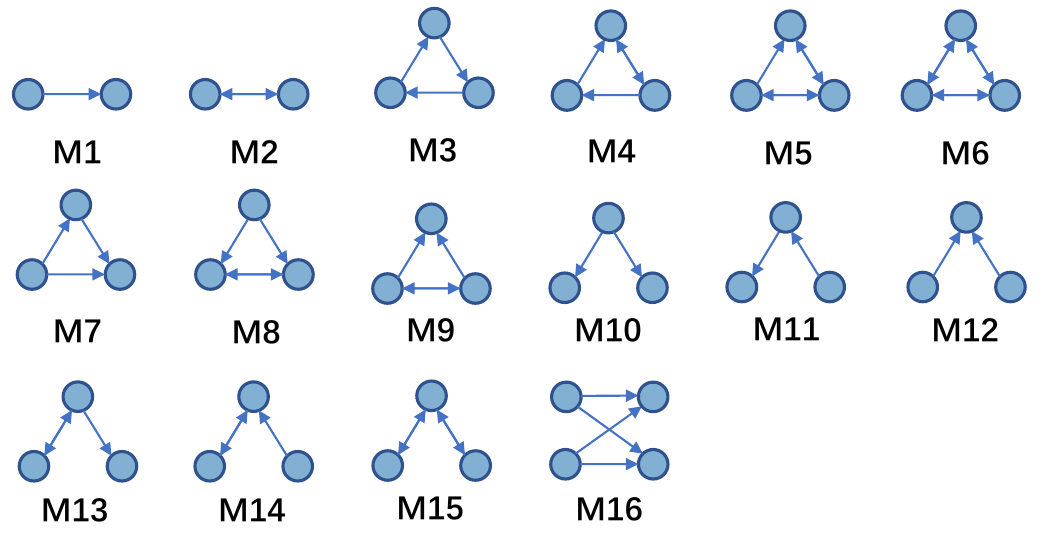}
    \caption{Directed network motifs.}
    \label{fig:motif}
\end{figure}

\textbf{Asset Transfer Motif Statistics:} Using the constructed asset transfer graph, we extract structural features to characterize the asset transfer patterns of each transaction. Network motifs, as higher-order network organizations, serve as effective tools for reflecting and extracting hidden information about network structures~\cite{benson2016higher}. Fig.~\ref{fig:motif} illustrates 16 directed network patterns composed of two and three vertices, along with a special four-vertex directed network pattern. These motif patterns have been extensively studied and shown to reveal unique properties of various networks. Inspired by MoTS~\cite{wu2023know} (a generalized semantic representation of transactions), we apply the concept of network motifs to characterize the asset transfer structure of transactions. For $tx_{h}$, we extract a 16-dimensional feature vector $v_h^{f t}$. The $i_{th}$ element of $v_h^{f t}$ indicates the frequency of the $i_{th}$ motif (as shown in Fig.~\ref{fig:motif}) within the constructed asset transfer relationship graph. We compute the directed motifs ${M1}$ to ${M16}$ through subgraph matrix calculations~\cite{wu2023know}, facilitating the semantic extraction of asset transfers for transactions.

\subsection{Message-passing Semantics}

In the \myModel~framework we designed, messaging semantic extraction focuses on elements of cross-chain message-passing. This section explores the encoding of transaction event logs that are closely associated with messaging. The differences in event logs triggered by cross-chain versus non-cross-chain behavior may yield unique textual features.

\begin{figure}[htbp]
    \centering
    \includegraphics[width=0.95\linewidth]{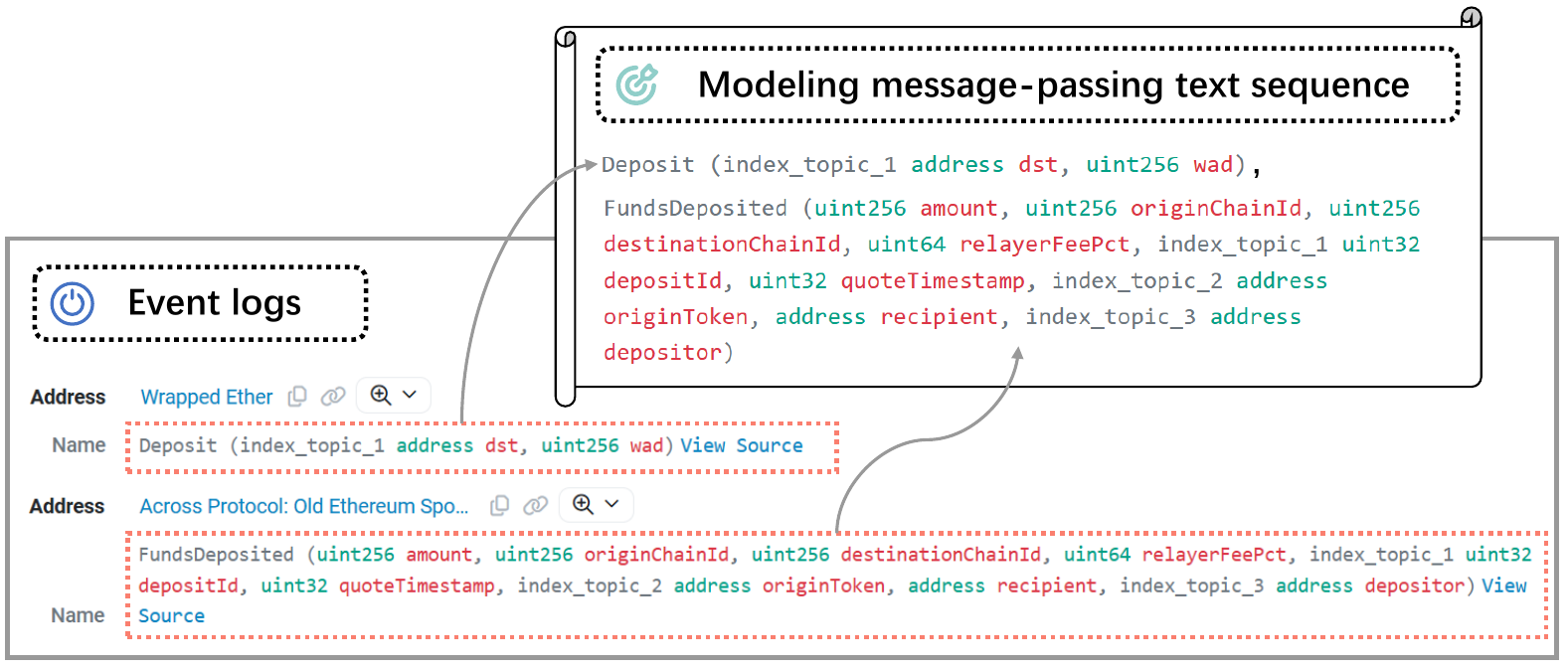}
    \caption{Message-passing Text Modeling.}
    \label{fig:event}
\end{figure}

\textbf{Message-passing Text Modeling:} A sequence of transaction event logs on a blockchain serves as an ordered record of various events and state changes occurring during transaction execution, reflecting the specific operations and outcomes involved. We utilize the event log sequence associated with $tx_{h}$ to construct a corresponding message-passing text sequence for each transaction. Specifically, we crawl all event log data related to the transaction through the RPC APIs provided by full blockchain nodes. We extract event names containing semantic information from each log and concatenate these names, punctuated appropriately, to form a narrative describing the transaction's execution process. To ensure the validity of the input sequence, we impose a maximum length limit, setting it to $max\_length=256$. If the text exceeds this limit, we perform truncation. As previously mentioned, we take the Ethereum network transaction 0x00f2 as an example to illustrate the process of extracting its message-passing text sequence, as shown in Fig.~\ref{fig:event}. This transaction includes two event logs, with event names ``Deposit" and ``FundsDeposited." Each event name includes the event title and the parameters involved in the operation (including parameter types and names).

\addtolength{\topmargin}{0.08in}
\textbf{Message-passing Event Encoding:} Using the constructed messaging text, we extract the corresponding textual features to characterize the execution process semantics of each transaction. Given that the format of event names resembles function declarations in programming languages, we employ pre-trained models such as CodeBERT~\cite{feng2020codebert} to obtain semantic representations of the text. These models, trained on large-scale datasets, provide robust initial parameters that enhance the efficiency and accuracy of subsequent tasks. For $tx_{h}$, we apply a multilayer perceptron (MLP) to refine the semantic embedding representations of events, ensuring alignment with the asset transfer semantic representation $v_h^{f t}$. The MLP's nonlinear transformation and feature extraction capabilities allow us to capture and integrate the complex semantic information of events effectively. This process involves multiple hidden layers that introduce nonlinearity through activation functions, enabling the model to learn deeper feature representations and thereby improve the performance of subsequent tasks. Ultimately, we obtain a 16-dimensional textual semantic representation of message-passing, denoted as $v_h^{m p}$.

\subsection{Semantic Integration}


After obtaining the semantic representation of asset transfer structures $v_h^{f t}$ and the semantic representation of message-passing text $v_h^{m p}$, we perform a vector concatenation operation on the two representations to obtain a complete semantic representation $v_h$ that incorporates both key cross-chain elements:

\vspace{-1ex}
\begin{equation}
v_h=\left[v_h^{f t}, v_h^{m p}\right]
\end{equation}

Subsequently, the complete semantic representation $v_h$ is input into the downstream classifier to facilitate the recognition of cross-chain transaction semantics.

\section{Experiments}

In this section, we evaluate the performance of the \myModel~framework by selecting 10 prominent cross-chain bridges from the Chainspot platform, which offers optimal cross-chain solutions for Web3 users and developers. We collect cross-chain transaction data from these bridges and conduct experiments. We provide an overview of the fundamental characteristics of the dataset, outline the evaluation metrics employed, assess the generality and generalization of the framework, and summarize our findings.

\subsection{Dataset}

Our dataset primarily consists of two components: cross-chain transaction labels and transaction information. The cross-chain transaction labels are derived from ten popular cross-chain bridge platforms, with detailed statistics on the number of transactions for each bridge presented in TABLE~\ref{tab:cct_volume_statistics}. The transaction information is sourced from three mainstream blockchain platforms: Ethereum (ETH), Binance Smart Chain (BSC), and Polygon (POL). We randomly sampled cross-chain transaction data from April 2021 to March 2024, which includes transactions initiated from ETH to be received by both BSC and POL. Additionally, we randomly sampled non-cross-chain transactions conducted on ETH, covering the period from March 2018 to June 2023. The statistics for non-cross-chain transactions, source chain deposit transactions, and target chain withdrawal transactions in the dataset are detailed in TABLE~\ref{tab:vol_tx}. It is noteworthy that the counts of deposit and withdrawal transactions are equal, as they both originate from the corresponding cross-chain transactions.


Regarding the labeling of cross-chain transactions, we provide further clarification here. First, we collected a substantial amount of transaction data from the blockchain platforms involved in the source chain over recent years. Based on these transaction records, we accessed the cross-chain bridge browsers and utilized their search functions to filter potential cross-chain transactions, subsequently scraping the search results to construct accurate labels for the cross-chain transactions.


\begin{table}
\renewcommand{\arraystretch}{1.6}
  \centering
  \caption{Cross-chain transaction volume statistics.}
  \scalebox{0.94}{
    \begin{tabular}{l|lll}
    \hline
    \multicolumn{1}{c|}{\multirow{2}[2]{*}{\textbf{\scalebox{1.2}{Bridge}}}} & \multicolumn{3}{c}{\textbf{\scalebox{1.2}{\# Transaction Pairs}}}                    \\ 
    \multicolumn{1}{c|}{}                                    & \centering \textbf{\scalebox{1.0}{\# ETH $\Rightarrow$ BSC}} & \multicolumn{1}{l|}{\centering \textbf{\scalebox{1.0}{\# ETH $\Rightarrow$ POL}}} & \textbf{\centering \scalebox{1.0}{\# All}} \\ \hline
    \scalebox{1.2}{Allbridge core}                     & \scalebox{1.2}{364}                     & \multicolumn{1}{l|}{\scalebox{1.2}{116}}               & \scalebox{1.2}{480}   \\
    \scalebox{1.2}{Celer cbridge}                      & \scalebox{1.2}{6,846}                   & \multicolumn{1}{l|}{\scalebox{1.2}{531}}               & \scalebox{1.2}{7,377} \\ 
    \scalebox{1.2}{Connext bridge}                     & \scalebox{1.2}{49}                      & \multicolumn{1}{l|}{\scalebox{1.2}{41}}                & \scalebox{1.2}{90}    \\
    \scalebox{1.2}{Multichain}                         & \scalebox{1.2}{1,655}                   & \multicolumn{1}{l|}{\scalebox{1.2}{664}}               & \scalebox{1.2}{2,319} \\ 
    \scalebox{1.2}{Polybridge}                         & \scalebox{1.2}{20}                      & \multicolumn{1}{l|}{\scalebox{1.2}{440}}               & \scalebox{1.2}{460}   \\
    \scalebox{1.2}{Stargate}                           & \scalebox{1.2}{716}                     & \multicolumn{1}{l|}{\scalebox{1.2}{125}}               & \scalebox{1.2}{841}   \\ 
    \scalebox{1.2}{Symbiosis}                          & \scalebox{1.2}{113}                     & \multicolumn{1}{l|}{\scalebox{1.2}{ 0}}                & \scalebox{1.2}{113}   \\
    \scalebox{1.2}{Synapse protocol}                   & \scalebox{1.2}{ 0}                      & \multicolumn{1}{l|}{\scalebox{1.2}{56}}                & \scalebox{1.2}{56}    \\ 
    \scalebox{1.2}{Transit swap}                       & \scalebox{1.2}{ 0}                      & \multicolumn{1}{l|}{\scalebox{1.2}{90}}                & \scalebox{1.2}{90}    \\
    \scalebox{1.2}{Wormhole}                           & \scalebox{1.2}{ 0}                     & \multicolumn{1}{l|}{\scalebox{1.2}{53}}                 & \scalebox{1.2}{53}    \\ \hline
    \textbf{\scalebox{1.2}{\# Total}}                                              & \textbf{\scalebox{1.2}{9,763}}                   & \multicolumn{1}{l|}{\textbf{\scalebox{1.2}{2,116}}}                   & \textbf{\scalebox{1.2}{11,879}}   \\ \hline
    \end{tabular}
  }
  \label{tab:cct_volume_statistics}%
\end{table}%


\begin{table}
\renewcommand{\arraystretch}{1.6}
  \centering
  \caption{Volume distribution of transactions.}
  \scalebox{1.05}{

  \begin{tabular}{lll}
    \hline
    \textbf{\scalebox{1.2}{\# Non-Cross-Chain}} & \textbf{\scalebox{1.2}{\# Deposit}}   & \textbf{\scalebox{1.2}{\# Withdrawal}} \\
    \textbf{\scalebox{1.2}{Transactions}}       & \textbf{\scalebox{1.2}{Transactions}} & \textbf{\scalebox{1.2}{Transactions}}  \\ \hline
    \scalebox{1.2}{10,183}             & \scalebox{1.2}{11,879}       & \scalebox{1.2}{11,879}
    \\ \hline
    \end{tabular}
  }
  \label{tab:vol_tx}%
\end{table}%

\subsection{Metrics}
We employ four metrics—precision, recall, accuracy, and F-score—to assess the performance of the model. Below is a brief description of each metric:

\vspace{-1ex}
\begin{equation}
\text { Precision }=\frac{\text { TP }}{\text { TP }+ \text { FP }}
\end{equation}

\begin{equation}
\text { Recall }=\frac{\text { TP }}{\text { TP }+ \text { FN }}
\end{equation}

\begin{equation}
\text { Accuracy }=\frac{\text { TP} + \text{TN}}{\text { TP } + \text{TN}  + \text { FP } + \text { FN }}
\end{equation}

\begin{equation}
\mathrm{F}-\text{score}=2 \times \frac{\text { Precision } \times \text { Recall }}{\text { Precision }+ \text { Recall }}
\end{equation}

\subsection{Baseline}
We explore variations of the pre-training module and classifier module within the proposed \myModel~framework. The pre-training module is implemented using CodeBERT~\cite{feng2020codebert}, GraphCodeBERT~\cite{guo2020graphcodebert}, and UniXcoder~\cite{guo2022unixcoder} as the pre-trained models. We select several machine learning models~\cite{pedregosa2011scikit} as classifiers, including Decision Tree (DT), Support Vector Machine (SVM), Random Forest (RF), and Adaptive Boosting (AdaBoost). The chosen pre-trained models and classifiers are combined in pairs to evaluate the overall performance of the \myModel~ framework. Additionally, we use the existing web3 general semantic method MoTS~\cite{wu2023know} as a baseline. To ensure a fair comparison, we pair MoTS with the classifiers utilized in this study and assess the predictive effectiveness of each method while maintaining consistency among the classifiers.

\subsection{Generality}


To verify the applicability of \myModel~in extracting semantics on mainstream cross-chain bridges, we conduct tests across all cross-chain bridges that we collected. We divide the ordinary transactions into training and test sets in an 80:20 ratio and similarly categorized both deposit and withdrawal transactions on each cross-chain bridge into training and test sets.

\begin{table}
\renewcommand{\arraystretch}{1.6}
  \centering
  \caption{Generality experiment results.}
  \scalebox{0.70}{
    \begin{tabular}{ll|llll}
    \hline
    \multicolumn{2}{c|}{\textbf{\scalebox{1.2}{Model}}}             & \textbf{\scalebox{1.2}{Precision}} & \textbf{\scalebox{1.2}{Recall}} & \textbf{\scalebox{1.2}{Accuracy}} & \textbf{\scalebox{1.2}{F1-macro}} \\ \hline
    \multicolumn{1}{l|}{\multirow{4}{*}{\scalebox{1.2}{DT}}}  & \scalebox{1.2}{MoTS}               & \scalebox{1.2}{{90.43\%}}        & \scalebox{1.2}{{90.43\%}}     & \scalebox{1.2}{{90.43\%}}       & \scalebox{1.2}{{90.83\%}}       \\
    \multicolumn{1}{l|}{}                     & \scalebox{1.2}{\myModel~(CodeBert)}      & \scalebox{1.2}{{99.09\%}}        & \scalebox{1.2}{{99.09\%}}     & \scalebox{1.2}{{99.09\%}}       & \scalebox{1.2}{{99.07\%}}       \\
    \multicolumn{1}{l|}{}                     & \scalebox{1.2}{\myModel~(GraphCodeBert)} & \scalebox{1.2}{{98.87\%}}        & \scalebox{1.2}{{98.87\%}}     & \scalebox{1.2}{{98.87\%}}       & \scalebox{1.2}{{98.84\%}}       \\
    \multicolumn{1}{l|}{}                     & \textbf{\scalebox{1.2}{\myModel~(Unixcoder)}}     & \textbf{\scalebox{1.2}{{99.72\%}}}        & \textbf{\scalebox{1.2}{{99.72\%}}}     & \textbf{\scalebox{1.2}{{99.72\%}}}       & \textbf{\scalebox{1.2}{{99.72\%}}}       \\ \hline
    \multicolumn{1}{l|}{\multirow{4}{*}{\scalebox{1.2}{SVM}}}  & \scalebox{1.2}{MoTS}               & \scalebox{1.2}{{90.25\%}}        & \scalebox{1.2}{{90.25\%}}     & \scalebox{1.2}{{90.25\%}}       & \scalebox{1.2}{{90.65\%}}       \\
    \multicolumn{1}{l|}{}                     & \scalebox{1.2}{\myModel~(CodeBert)}      & \scalebox{1.2}{{96.89\%}}        & \scalebox{1.2}{{96.89\%}}     & \scalebox{1.2}{{96.89\%}}       & \scalebox{1.2}{{96.84\%}}       \\
    \multicolumn{1}{l|}{}                     & \scalebox{1.2}{\myModel~(GraphCodeBert)} & \scalebox{1.2}{{99.48\%}}        & \scalebox{1.2}{{99.48\%}}     & \scalebox{1.2}{{99.48\%}}       & \scalebox{1.2}{{99.48\%}}       \\
    \multicolumn{1}{l|}{}                     & \textbf{\scalebox{1.2}{\myModel~(Unixcoder)}}     & \textbf{\scalebox{1.2}{{99.57\%}}}        & \textbf{\scalebox{1.2}{{99.57\%}}}     & \textbf{\scalebox{1.2}{{99.57\%}}}       & \textbf{\scalebox{1.2}{{99.56\%}}}       \\ \hline
    \multicolumn{1}{l|}{\multirow{4}{*}{\scalebox{1.2}{RF}}}  & \scalebox{1.2}{MoTS}               & \scalebox{1.2}{{90.43\%}}        & \scalebox{1.2}{{90.43\%}}     & \scalebox{1.2}{{90.43\%}}       & \scalebox{1.2}{{90.83\%}}       \\
    \multicolumn{1}{l|}{}                     & \scalebox{1.2}{\myModel~(CodeBert)}      & \scalebox{1.2}{{98.72\%}}        & \scalebox{1.2}{{98.72\%}}     & \scalebox{1.2}{{98.72\%}}       & \scalebox{1.2}{{98.68\%}}       \\
    \multicolumn{1}{l|}{}                     & \scalebox{1.2}{\myModel~(GraphCodeBert)} & \scalebox{1.2}{{98.81\%}}        & \scalebox{1.2}{{98.81\%}}     & \scalebox{1.2}{{98.81\%}}       & \scalebox{1.2}{{98.78\%}}       \\
    \multicolumn{1}{l|}{}                     & \textbf{\scalebox{1.2}{\myModel~(Unixcoder)}}     & \textbf{\scalebox{1.2}{{99.37\%}}}        & \textbf{\scalebox{1.2}{{99.37\%}}}     & \textbf{\scalebox{1.2}{{99.37\%}}}       & \textbf{\scalebox{1.2}{{99.35\%}}}       \\ \hline
   \multicolumn{1}{l|}{\multirow{4}{*}{\scalebox{1.2}{AdaBoost}}}  & \scalebox{1.2}{MoTS}               & \scalebox{1.2}{{76.60\%}}        & \scalebox{1.2}{{76.60\%}}     & \scalebox{1.2}{{76.60\%}}       & \scalebox{1.2}{{76.62\%}}       \\
    \multicolumn{1}{l|}{}                     & \scalebox{1.2}{\myModel~(CodeBert)}      & \scalebox{1.2}{{94.88\%}}        & \scalebox{1.2}{{94.88\%}}     & \scalebox{1.2}{{94.88\%}}       & \scalebox{1.2}{{94.79\%}}       \\
    \multicolumn{1}{l|}{}                     & \scalebox{1.2}{\myModel~(GraphCodeBert)} & \scalebox{1.2}{{98.40\%}}        & \scalebox{1.2}{{98.40\%}}     & \scalebox{1.2}{{98.40\%}}       & \scalebox{1.2}{{98.37\%}}       \\
    \multicolumn{1}{l|}{}                     & \textbf{\scalebox{1.2}{\myModel~(Unixcoder)}}     & \textbf{\scalebox{1.2}{{99.65\%}}}        & \textbf{\scalebox{1.2}{{99.65\%}}}     & \textbf{\scalebox{1.2}{{99.65\%}}}       & \textbf{\scalebox{1.2}{{99.64\%}}}       \\ \hline
    \end{tabular}
  }
  \label{tab:generality_result}%
\end{table}%

The classification results of various models are presented in TABLE~\ref{tab:generality_result}. The comparison indicates that all variants of the proposed framework outperform existing semantic extraction methods, with a prediction accuracy of up to {99.72\%} achieved when the pre-training model of the \myModel~framework is set to UniXcoder and the classifier is set to Decision Tree (DT). Furthermore, the performance of the \myModel~framework across all metrics remains above {94\%}, demonstrating its strong generalization capability. The variation in prediction performance among the different variants does not exceed {5\%}, suggesting that the framework's generalization is minimally affected by the choice of pre-training model and classifier. In summary, when classifiers are held constant, the selection of UniXcoder as the pre-training model yields the best results. Notably, even with the simplest classifier, DT, the framework effectively distinguishes between different transactions, indicating significant semantic differences among transactions within the high-dimensional semantic vector space.

\subsection{Generalizability}


To verify the ability of \myModel~to extract semantics on unseen new cross-chain bridges, we select four cross-chain bridges with the highest transaction volumes from a total of ten as our training subjects: Allbridge Core, Celer cBridge, Multichain, and Stargate. Subsequently, we conduct tests on the remaining six cross-chain bridges. Additionally, we divide the ordinary transactions into training and testing sets in an 80:20 ratio.

\begin{table}
\renewcommand{\arraystretch}{1.6}
  \centering
  \caption{Generalizability experiment results.}
  \scalebox{0.68}{
    \begin{tabular}{ll|llll}
    \hline
    \multicolumn{2}{c|}{\textbf{\scalebox{1.2}{Model}}}             & \textbf{\scalebox{1.2}{Precision}} & \textbf{\scalebox{1.2}{Recall}} & \textbf{\scalebox{1.2}{Accuracy}} & \textbf{\scalebox{1.2}{F1-macro}} \\ \hline
    \multicolumn{1}{l|}{\multirow{4}{*}{\scalebox{1.2}{DT}}}  & \scalebox{1.2}{MoTS}               & \scalebox{1.2}{{78.03\%}}        & \scalebox{1.2}{{78.03\%}}     & \scalebox{1.2}{{78.03\%}}       & \scalebox{1.2}{{77.03\%}}       \\
    \multicolumn{1}{l|}{}                     & \scalebox{1.2}{\myModel~(CodeBert)}      & \scalebox{1.2}{{88.06\%}}        & \scalebox{1.2}{{88.06\%}}     & \scalebox{1.2}{{88.06\%}}       & \scalebox{1.2}{{87.41\%}}       \\
    \multicolumn{1}{l|}{}                     & \textbf{\scalebox{1.2}{\myModel~(GraphCodeBert)}} & \textbf{\scalebox{1.2}{{93.36\%}}}        & \textbf{\scalebox{1.2}{{93.36\%}}}     & \textbf{\scalebox{1.2}{{93.36\%}}}       & \textbf{\scalebox{1.2}{{93.48\%}}}       \\
    \multicolumn{1}{l|}{}                     & \scalebox{1.2}{\myModel~(Unixcoder)}     & \scalebox{1.2}{{86.10\%}}        & \scalebox{1.2}{{86.10\%}}     & \scalebox{1.2}{{86.10\%}}       & \scalebox{1.2}{{84.73\%}}       \\ \hline
    \multicolumn{1}{l|}{\multirow{4}{*}{\scalebox{1.2}{SVM}}}  & \scalebox{1.2}{MoTS}               & \scalebox{1.2}{{84.78\%}}        & \scalebox{1.2}{{84.78\%}}     & \scalebox{1.2}{{84.78\%}}       & \scalebox{1.2}{{83.82\%}}       \\
    \multicolumn{1}{l|}{}                     & \scalebox{1.2}{\myModel~(CodeBert)}      & \scalebox{1.2}{{85.37\%}}        & \scalebox{1.2}{{85.37\%}}     & \scalebox{1.2}{{85.37\%}}       & \scalebox{1.2}{{84.69\%}}       \\
    \multicolumn{1}{l|}{}                     & \textbf{\scalebox{1.2}{\myModel~(GraphCodeBert)}} & \textbf{\scalebox{1.2}{{94.81\%}}}        & \textbf{\scalebox{1.2}{{94.81\%}}}     & \textbf{\scalebox{1.2}{{94.81\%}}}       & \textbf{\scalebox{1.2}{{94.85\%}}}       \\
    \multicolumn{1}{l|}{}                     & \scalebox{1.2}{\myModel~(Unixcoder)}     & \scalebox{1.2}{{94.40\%}}        & \scalebox{1.2}{{94.40\%}}     & \scalebox{1.2}{{94.40\%}}       & \scalebox{1.2}{{94.50\%}}       \\ \hline
    \multicolumn{1}{l|}{\multirow{4}{*}{\scalebox{1.2}{RF}}}  & \scalebox{1.2}{MoTS}               & \scalebox{1.2}{{79.30\%}}        & \scalebox{1.2}{{79.30\%}}     & \scalebox{1.2}{{79.30\%}}       & \scalebox{1.2}{{78.37\%}}       \\
    \multicolumn{1}{l|}{}                     & \scalebox{1.2}{\myModel~(CodeBert)}      & \scalebox{1.2}{{85.05\%}}        & \scalebox{1.2}{{85.05\%}}     & \scalebox{1.2}{{85.05\%}}       & \scalebox{1.2}{{84.07\%}}       \\
    \multicolumn{1}{l|}{}                     & \textbf{\scalebox{1.2}{\myModel~(GraphCodeBert)}} & \textbf{\scalebox{1.2}{{88.52\%}}}        & \textbf{\scalebox{1.2}{{88.52\%}}}     & \textbf{\scalebox{1.2}{{88.52\%}}}       & \textbf{\scalebox{1.2}{{87.78\%}}}       \\
    \multicolumn{1}{l|}{}                     & \scalebox{1.2}{\myModel~(Unixcoder)}     & \scalebox{1.2}{{85.25\%}}        & \scalebox{1.2}{{85.25\%}}     & \scalebox{1.2}{{85.25\%}}       & \scalebox{1.2}{{83.93\%}}       \\ \hline
   \multicolumn{1}{l|}{\multirow{4}{*}{\scalebox{1.2}{AdaBoost}}}  & \scalebox{1.2}{MoTS}               & \scalebox{1.2}{{81.11\%}}        & \scalebox{1.2}{{81.11\%}}     & \scalebox{1.2}{{81.11\%}}       & \scalebox{1.2}{{81.11\%}}       \\
    \multicolumn{1}{l|}{}                     & \scalebox{1.2}{\myModel~(CodeBert)}      & \scalebox{1.2}{{85.84\%}}        & \scalebox{1.2}{{85.84\%}}     & \scalebox{1.2}{{85.84\%}}       & \scalebox{1.2}{{85.48\%}}       \\
    \multicolumn{1}{l|}{}                     & \textbf{\scalebox{1.2}{\myModel~(GraphCodeBert)}} & \textbf{\scalebox{1.2}{{93.12\%}}}        & \textbf{\scalebox{1.2}{{93.12\%}}}     & \textbf{\scalebox{1.2}{{93.12\%}}}       & \textbf{\scalebox{1.2}{{93.01\%}}}       \\
    \multicolumn{1}{l|}{}                     & \scalebox{1.2}{\myModel~(Unixcoder)}     & \scalebox{1.2}{{83.71\%}}        & \scalebox{1.2}{{83.71\%}}     & \scalebox{1.2}{{83.71\%}}       & \scalebox{1.2}{{82.27\%}}       \\ \hline
    \end{tabular}
  }
  \label{tab:generalizability_result}%
\end{table}%

The classification results of various models are presented in TABLE~\ref{tab:generalizability_result}. The comparison indicates that all variants of the proposed framework outperform existing semantic extraction methods. Notably, when the pre-training model of the \myModel~framework is set to GraphCodeBERT and the classifier to Support Vector Machine (SVM), we achieved a superior prediction accuracy of {94.81\%}, demonstrating the framework's strong generalization capability. In summary, the selection of GraphCodeBERT as the pre-training model yields the best results in effectively predicting unseen cross-chain bridge samples, while maintaining consistency among the classifiers.

\subsection{Semantic Analysis}


To further investigate the factors influencing the semantic extraction of cross-chain transactions, we analyze the \myModel~framework variant that yields the best prediction results in the generalizability experiments. This analysis focuses on the transaction graph structures and event log texts for each type of transaction in the prediction results.

\begin{figure}[htbp]
    \centering
    \includegraphics[width=1\linewidth]{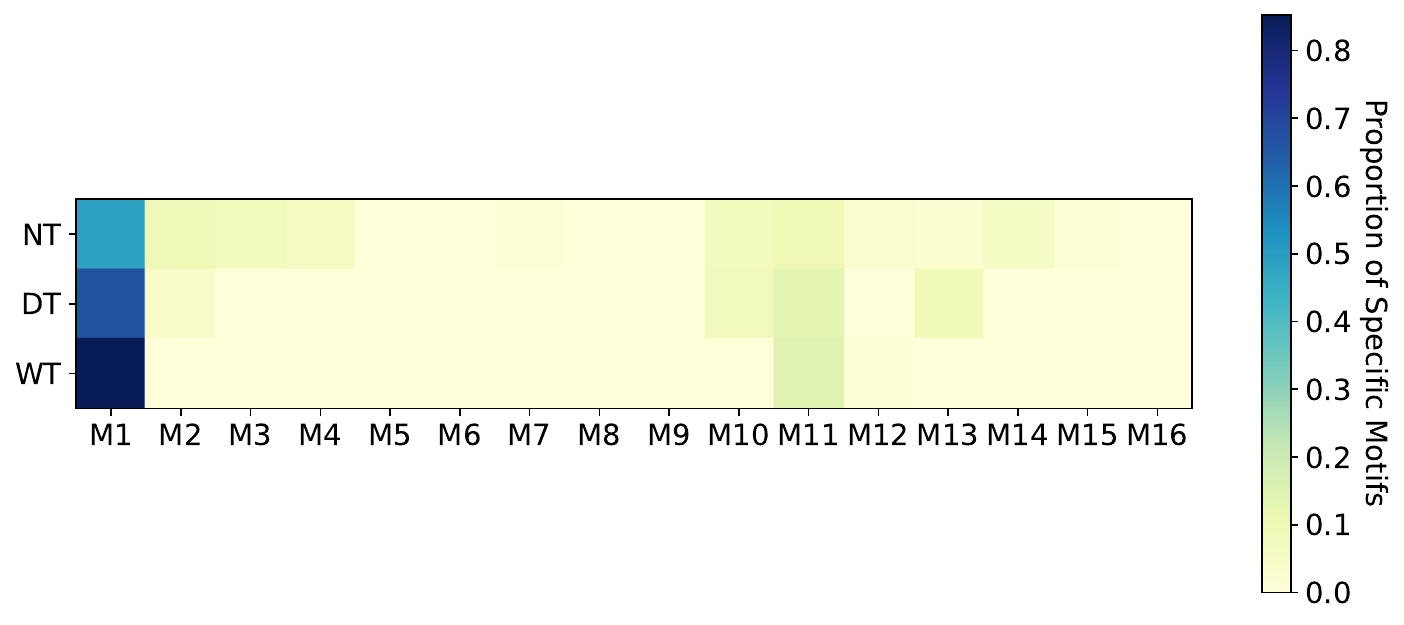}
    \caption{Heatmap of Motif Statistics.}
    \label{fig:motif_analyze}
\end{figure}

\textbf{Asset Transfer Semantics:} We examine the differences in the distribution of motifs between cross-chain and non-cross-chain transactions. Utilizing heat maps, we quantify the 16 types of motifs for both transaction categories, as illustrated in Fig.~\ref{fig:motif_analyze}. Given that higher-order motifs typically encompass lower-order ones, our analysis emphasize the distribution differences of higher-order motifs. Our findings indicate that: 1) Non-cross-chain transactions involve a broader range of motifs compared to cross-chain transactions; 2) Cross-chain deposit transactions primarily concentrate on motifs 10, 11, and 13; and 3) Cross-chain withdrawal transactions predominantly feature motif 11. These results suggest that the distribution of network motifs in cross-chain transactions is more focused, exhibiting notable structural uniqueness.

\begin{figure}[htbp]
    \centering
    \subfigure[Non-Cross-Chain Transactions]{%
        \centering
        \includegraphics[width=0.64\linewidth]{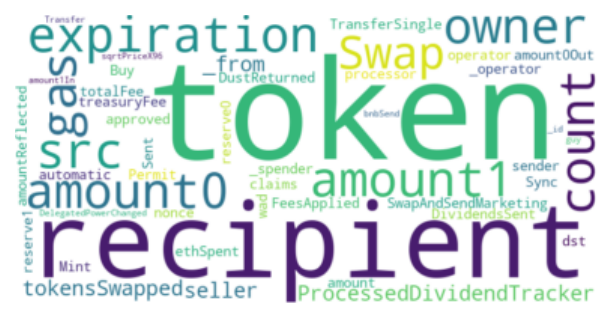}
    }
    \subfigure[Deposit Transactions]{%
        \centering
        \includegraphics[width=0.45\linewidth]{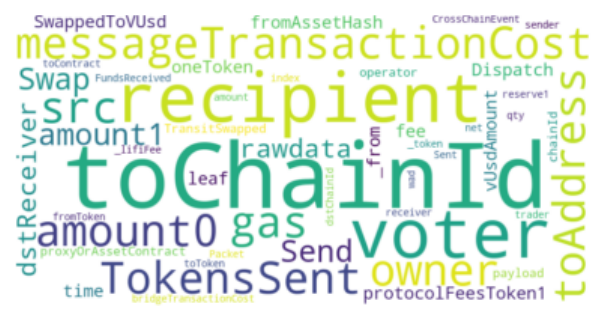}
    }
    \subfigure[Withdrawal Transactions]{%
        \centering
        \includegraphics[width=0.45\linewidth]{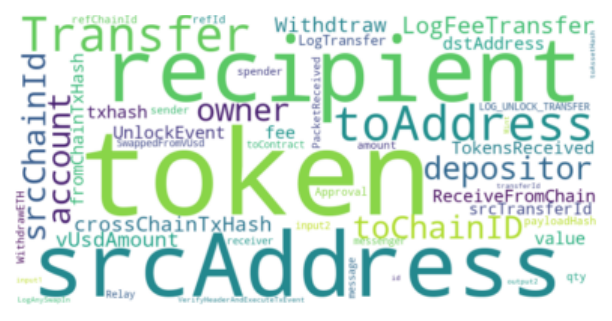}
    }
    \caption{Word clouds of different types of transactions.}
    \label{fig:event_analyze}
\end{figure}

\textbf{Message-passing Semantics:} We also explore the distribution of events associated with cross-chain transactions compared to non-cross-chain transactions. We extract event names and their corresponding parameter names for both types of transactions, calculating the frequency of each term, as depicted in the event cloud diagram in Fig.~\ref{fig:event_analyze}. Our observations reveal that: 1) Common terms such as ``token" and ``permit" appear more frequently in the event logs of non-cross-chain transactions; 2) Cross-chain transactions contain a higher prevalence of terms with ``Id," such as ``transferId," ``toChainId," and ``mintId"; and 3) The term ``toChainId" is particularly prevalent in cross-chain deposit transactions, while ``mintId" and ``toAssetHash" are more frequent in cross-chain withdrawal transactions. This indicates that the event logs of cross-chain transactions typically incorporate proprietary terminology related to cross-chain mechanisms, showcasing distinct textual characteristics compared to non-cross-chain transactions.

\section{Conclusions}

In this paper, we investigate transaction semantic extraction within blockchain cross-chain scenarios and propose a cross-chain semantic extraction framework named \myModel, which comprises two core modules: asset transfer semantic extraction and message-passing semantic extraction. Unlike existing methods, this framework is specifically designed for cross-chain scenarios. It models token transfers as low-level semantic networks and constructs message-passing events as text sequences to extract semantics. Ultimately, it integrates these two inputs into a downstream classifier to categorize cross-chain deposits, cross-chain withdrawals, and non-cross-chain transactions. For data collection, we construct a validation dataset consisting of 11,879 cross-chain transactions and 10,183 non-cross-chain transactions. Experimental results demonstrate that the framework achieves prediction accuracies of {99\%} and over {94\%} in terms of generality and generalization, respectively. Additionally, the analysis reveals significant differences in asset transfer patterns and event word frequencies between cross-chain and non-cross-chain transactions. While the method relies heavily on existing pre-trained models for event log processing, future research will focus on enhancing the semantic representation of cross-chain transactions through model fine-tuning and advanced modeling techniques. Overall, this paper offers new insights into the coexistence of multiple blockchains and the dynamics of cross-chain ecosystems.

\section*{Acknowledgment}


The work described in this paper is supported by the National Key Research and Development Program of China (2023YFB2704700), the National Natural Science Foundation of China (62372485, 623B2102, and 62332004), the Natural Science Foundations of Guangdong Province (2023A1515011314), the Department of Education of Guangdong Province (2024ZDZX1001), the Fundamental Research Funds for the Central Universities of China, Sun Yat-sen University (24lgqb018), and Shanghai Committee of Science and Technology, China (23511101000). 

\bibliographystyle{IEEEtran}
\bibliography{reference.bib}
\end{document}